\documentclass[preprint,prd,tightenlines,nofootinbib,superscriptaddress]{revtex4}

\usepackage{color}
\usepackage{graphicx}
\definecolor{red}{rgb}{1,0,0}
\def\lesssim{\ \hbox{\raise 2pt \hbox{$<$} \kern -13pt
                     \lower 3pt \hbox{$\sim$}}\ }
\def\greatersim{\ \hbox{\raise 2pt \hbox{$>$} \kern -13pt
                     \lower 3pt \hbox{$\sim$}}\ }
\def\herwig{{\sc Herwig}}
\def\pythia{{\sc Pythia}}
\input epsf.tex
\def\desepsf(#1 width #2){\epsfxsize=#2 \epsfbox{#1}}

\usepackage{amsmath,bm}
\begin{document}

\hspace*{11.9 cm} {\small DESY-09-119} 

\hspace*{11.9 cm} {\small OUTP-09-15-P}

\vspace*{1 cm} 

\title{Forward Jet  Production at the Large Hadron Collider}
\author{M.\ Deak}
\affiliation{Deutsches Elektronen Synchrotron, D-22603 Hamburg}
\author{F.\ Hautmann} 
\affiliation{Theoretical Physics, 
University of Oxford,    Oxford OX1 3NP}
\author{H.\ Jung}
\affiliation{Deutsches Elektronen Synchrotron, D-22603 Hamburg}
\author{K.\ Kutak}
\affiliation{Deutsches Elektronen Synchrotron, D-22603 Hamburg}

\begin{abstract}
 At   the Large Hadron Collider  (LHC) it will become  possible   
for the first time      to 
investigate experimentally 
 the forward region     in hadron-hadron collisions  
   via  high-$p_T$ processes. 
 In the LHC forward kinematics     
   QCD  logarithmic  corrections in the 
hard transverse momentum  and   
in the large rapidity  interval  may both  be 
 quantitatively   significant.   
 We analyze   the  hadroproduction  of forward jets in the framework of 
QCD  high-energy factorization,   which 
allows  one  to 
 resum consistently both kinds of corrections to higher orders in QCD perturbation theory.   
We     compute  the  
short-distance matrix elements  needed to evaluate the 
factorization formula  at  fully  exclusive  level.   
   We 
discuss  numerically    dynamical features  of  multi-gluon emission at   large 
angle   encoded 
in the factorizing   high-energy amplitudes.  
\end{abstract}


\maketitle

\section{Introduction}
\label{sec:intro}

Experiments at the  Large Hadron Collider  (LHC) 
  will  explore  the  region of  large    rapidities  
 both with   general-purpose 
 detectors and with dedicated instrumentation, including 
  forward  calorimeters   and proton 
  taggers~\cite{cmsfwd,atlasfwd,fp420,cmstotem,aslano,grothe,heralhc}.     
The    forward-physics   program 
involves   a wide range of topics, from  
new particle discovery  processes~\cite{fp420,fwdhiggs,fwdmssm}   
to  new  aspects of   
strong interaction physics~\cite{heralhc,denterria,cerci}  to 
heavy-ion  collisions~\cite{accardi03,heavy-ion-cmsnote}. 
Owing to  the  large center-of-mass energy    and 
  the unprecedented experimental coverage at large rapidities,  
 it becomes possible for the first time to investigate   the  forward region     
 with  high-$p_T$ probes.

\begin{figure}[htb]
\vspace{45mm}
\includegraphics{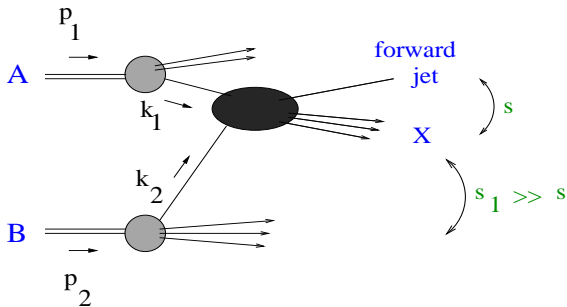}
\caption{Jet  production in the forward rapidity region 
in  hadron-hadron collisions.} 
\label{fig:forwpicture}
\end{figure}

\begin{figure}[htb]
\vspace{15mm}
\includegraphics{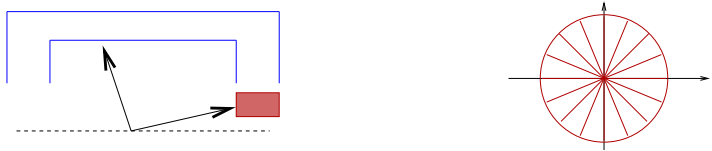}
\caption{(Left)    High-$p_T$  events  in the 
forward and central detectors; (right) azimuthal plane segmentation.} 
\label{fig:azimcorr}
\end{figure}

 The hadroproduction of a  forward jet   associated  
with  
  a   hard final state $X$  is pictured 
in Fig.~\ref{fig:forwpicture}.    
The kinematics 
  of the process     is  characterized 
by the  large  ratio  of sub-energies  $s_1  /  s \gg 1 $   
 and  highly asymmetric longitudinal momenta in the partonic initial 
  state, \mbox{$ k_1 \cdot p_2  \gg    k_2 \cdot p_1$}.  
 At the LHC the use of  forward calorimeters  allows  one to  
  measure    events  where   jet transverse momenta 
  $p_T  >   20$ GeV   are produced  several units of rapidity 
  apart,  $\Delta y   \greatersim 4 \div 6$~\cite{cmsfwd,aslano,heralhc}.  
 Working at    polar angles that are  small   but   sufficiently  far  from the beam axis 
 not to be affected by  beam remnants,     one measures 
 azimuthal plane correlations  
  between   high-$p_T$  events  
  widely  separated    in rapidity      (Fig.~\ref{fig:azimcorr}).

The   presence of multiple   large-momentum scales  
implies   that, as   was    recognized in~\cite{muenav,vddtang,stirl94},    
reliable  theoretical predictions   for forward jets 
can only be obtained after  summing  
 logarithmic  QCD corrections at high energy
 to all orders in $\alpha_s$.  
This   has  motivated    efforts~\cite{webetal99,orrsti,stirvdd,andsab}  to 
 construct   new   
  algorithms  for   Monte Carlo  event  generators capable of 
   describing   jet   production    beyond  the central rapidity region. 
 Note that an  analogous observation applies to  
  forward jets  associated 
 to  deeply    inelastic  scattering~\cite{mueproc90c,forwdis92}.    
 Indeed, measurements of 
 forward jet cross sections at  HERA~\cite{heraforw}  indicate  that 
 neither fixed-order next-to-leading  
 calculations  nor  standard shower 
 Monte Carlo   
     generators~\cite{webetal99,heraforw,web95}, e.g.   \pythia\ 
  or \herwig,  are    able to   describe 
 forward jet  $ep$ data.
  Improved methods to evaluate QCD    
     predictions 
  are needed to treat the multi-scale 
 region implied by  the  forward kinematics. 
 
In this   work  we  move on  from the observation  that    
  realistic  phenomenology  in the LHC forward region 
will   require   taking  into  account    at   higher  order     
 both logarithmic corrections   in  the large  rapidity  interval  
(of  high-energy type)  
and logarithmic corrections  in  
the hard transverse momentum (of collinear type).  
The theoretical framework to resum   consistently 
both  kinds of logarithmic corrections  in QCD  
 perturbation theory    is based on  high-energy    factorization at 
fixed transverse momentum~\cite{hef}.

This formulation    depends  on  unintegrated distributions for 
parton splitting, obeying appropriate evolution equations, and short-distance, 
process-dependent matrix elements.   
The  unintegrated-level evolution 
is given by  evolution equations in rapidity,   or angle,  parameters.   
Different  forms of the  evolution, valid  in   different kinematic regions, 
are available.    See~\cite{jcc-lc08,fhau_pol,ted,fhfeb07}, and 
references therein, for recent 
 work  in this area 
  and reviews. 
The   short-distance matrix elements,    needed  in the  evaluation of  the 
   factorization formula,    are the subject of this paper.  
We obtain their explicit expressions     
   in a fully exclusive form,   including  all partonic channels,  
   and present   results   of numerically integrating them over final states.  
   Such matrix elements,  though not    on shell,   are 
 gauge invariant and   perturbatively calculable.  They  factorize in the high energy  
 limit 
   in front of (unintegrated)  distributions for parton splitting not only in the collinear  
   emission region but also at finite angle.  In particular,     
they   can  serve to take  
into 
    account    effects of coherence from multi-gluon emission,  away 
from small angles,  
which become important for correlations among jets 
  across long      separations in rapidity.     
   We  give a numerical   illustration  of the high-$k_T$  behavior resulting 
   from such finite-angle radiation.

 On one hand, 
 once convoluted with the small-x  gluon Green's function 
 according to the 
 method~\cite{hef,ch94}, these matrix elements 
  control the summation of high-energy 
 logarithmic corrections, contributing    both 
 to the   next-to-leading-order BFKL kernel~\cite{fadlip98} 
 and to the jet impact factors~\cite{mc98,schw0703}.  
   On  the other hand, they can   be used in a shower 
   Monte Carlo     implementing  parton-branching kernels 
   at unintegrated level (see e.g.~\cite{jadach09,hj_ang}  for recent works)  
   to   generate fully exclusive events.   We leave these applications to 
 a   separate  paper.

The paper is organized   as follows.  
After   recalling the    factorized form of the cross sections 
in Sec.~\ref{sec2}, we  present    the high-energy amplitudes in Sec.~\ref{sec3}, and 
discuss     basic  properties  and numerical results in   Sec.~\ref{sec4}. We 
summarize   in Sec.~\ref{sec:concl}.

\section{High-energy factorized cross sections}
\label{sec2}

\begin{figure}[htb]
\vspace{44mm}
\includegraphics{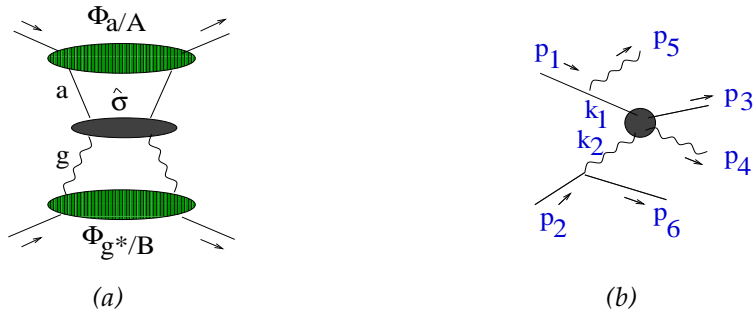}
\caption{(a) Factorized structure of the cross section; (b) a  graph contributing  to the 
$q g$ channel matrix element.} 
\label{fig:sec2}
\end{figure}

High-energy factorization~\cite{hef} allows one to decompose the cross section 
for the process of   Fig.~\ref{fig:forwpicture}  into  partonic distributions (in  general, 
unintegrated) and 
hard-scattering   kernels,  obtained  via the  high-energy 
projectors~\cite{hef,ch94}  from   the amplitudes for  the process 
$p_1 +  p_2 \to p_3 + p_4 + 2$ massless partons. 
The basic structure is depicted in Fig.~\ref{fig:sec2}.

With reference to the notation of   Fig.~\ref{fig:sec2}, 
let  us   work in the center of mass frame of  the incoming momenta 
\begin{equation}
\label{kinep1p2}
p_1 = \sqrt{S / 2}  ( 1 , 0, 0_{T} ) \;\;\;, \;\;\;\;\;\;  \;\;\;\;\;\; 
 p_2 = \sqrt{S / 2}  (  0, 1, 0_{T} )  
 \;\;\;, \;\;\;\;\;\;  \;\;\;\;\;\;   2 p_1 \cdot p_2 = S  \;\;\;,  
\end{equation}
where,  for   any  four-vector,      $p^\mu = ( p^+, p^- , p_{T}) $, with    
 $p^\pm = (p^0 \pm p^3 ) / \sqrt{2}$ and $p_{T}$ two-dimensional euclidean 
 vector.   
Let us  parameterize the exchanged momenta   in terms of     
purely transverse four-vectors $k_{\perp }$  and $k_{\perp 1}$ and 
longitudinal 
momentum  fractions $\xi_i$ and $ {\overline \xi}_i$    as  
\begin{equation}
\label{kinek1k2}
p_1 -  p_5 =   k_1 =   \xi_1  p_1 + k_{\perp 1} + {\overline \xi}_1 p_2  \;\;, \;\;\; 
p_2 -  p_6 =   k_2 =   \xi_2  p_2 + k_{\perp } + {\overline \xi}_2 p_1 \;\; . 
\end{equation}
For  high energies   we  can  introduce   strong  
ordering in the   longitudinal momenta, $\xi_1 \gg   |  {\overline \xi}_2 |$,   
  $\xi_2 \gg   |  {\overline \xi}_1 |$.  Further,  we   make the  forward region 
approximations $(p_4+ p_6)^2  \gg  (p_3 +p_4)^2   $, $k_1 \simeq   \xi_1  p_1$, 
$k_2 \simeq    \xi_2  p_2 + k_{\perp }$, so that 
\begin{equation}
\label{fwdkin}
p_5 \simeq  (1 - \xi_1 ) p_1 \;\;\;   ,  \;\;\;\;\;  p_6 \simeq   (1 - \xi_2 ) p_2 - k_\perp   
 \;\;\;   ,  \;\;\;\;\;  
\xi_1 \gg \xi_2     \;\;  .  
\end{equation}

It is convenient to  define  the rapidity-weighted average  of  dijet transverse 
momenta,  
\begin{equation}
\label{qtdef}
Q_T = (1-\nu) p_{T  4} - \nu p_{T  3}  \;\;, \;\;\;    {\rm{where}}  \;\;\; 
\nu =  (p_2 \, p_4) / [(p_2 \, p_1) -  (p_2 \, p_5)]   \;\;,  
\end{equation}
and  the azimuthal angle 
\begin{equation}
\label{phidef}
\cos \varphi = Q_T \cdot k_T / |  Q_T |  | k_T |   \;\; . 
\end{equation}
We consider the differential     jet  cross  section 
in $Q_T$  and   $\varphi$. 

    According to the  factorization~\cite{hef,mc98},   the  jet  cross  section  
can be computed  as    (Fig.~\ref{fig:sec2}a)
\begin{equation}
\label{forwsigma}
   {{d   \sigma  } \over 
{ d Q_T^2 d \varphi}} =  \sum_a  \int  \     
d \xi_1 \ d \xi_2  \  d^2 k_T \ 
 \phi_{a/A} (\xi_1)  \    {{d   {\widehat  \sigma}   } \over 
{ d Q_T^2 d \varphi  }}    ( \xi_1 \xi_2 S ,   k_T , Q_T , \varphi )  \ 
\phi_{g^*/B}  (\xi_2,   k_T)     \;\; , 
\end{equation}
where  the sum goes over  parton  species, $\phi$ are the parton distributions  
defined from  the unintegrated 
Green's functions introduced in~\cite{ch94} for both gluon and quark cases,  and 
$ {\widehat  \sigma} $  is the  hard   cross section,  calculable 
from  the high-energy limit  of  perturbative 
amplitudes  (Fig.~\ref{fig:sec2}b). 

The  physical picture underlying   Eq.~(\ref{forwsigma})    is  based on the  fact  
that 
initial-state parton configurations  contributing to  
forward production are asymmetric, 
with the parton in the top subgraph being  probed near  the mass shell and  
large  $ x $,  
while  the parton in  the bottom subgraph is off-shell and small-$x$.
 Eq.~(\ref{forwsigma}) embodies this picture through the 
 longitudinal and transverse momentum dependences of  both 
  $\phi$  and $ {\widehat  \sigma} $. 

For phenomenological studies 
we  will be interested in  coupling  Eq.~(\ref{forwsigma})  to  parton showers  
to achieve a full description of the  associated final states. 
To this end we need  
the matrix elements defining the  hard-scattering kernels          
  in a fully exclusive form.    We  
 give the results in the next section.

\section{Matrix elements for fully exclusive events}
\label{sec3}

The matrix elements determining  the hard-scattering kernels  
$ {\widehat  \sigma} $ 
can be  viewed  as  a suitably defined  off-shell continuation of 
scattering   amplitudes at lower order~\cite{hef}.  They can be obtained by applying  to   
scattering amplitudes    $ \mathcal{M}$   the high-energy eikonal 
projectors~\cite{hef,ch94},  
\begin{equation}
\label{proj}
 \mathcal{M}^H  =    P^{H \  \mu_1 \mu_2 ... }_{(eik)} \   \mathcal{M}_{ \mu_1 \mu_2  ...}  
 (k_1 , k_2, \{ p_i \} )      \;\; , 
 \;\; \; 
  P^{H\  \mu_1 \mu_2 ... }_{(eik)} 
   \propto  { { 2 k^{\mu_1}_{\perp 1} k^{\mu_2}_{\perp 2} } \over { \sqrt{   
k^{2}_{\perp 1} k^{2}_{\perp 2} }  }} \;\; . 
\end{equation}
Although they  are not evaluated  on shell, 
they are gauge invariant and their expressions are  simple.  
The utility of these  matrix elements is that  in the high-energy limit 
they factorize not only in the collinear emission region but also in the large-angle 
emission region. 
As long as 
 the factorization is carried out  in terms of distributions for parton splitting 
  at  fixed transverse momentum, as in   Eq.~(\ref{forwsigma}), 
they  can be useful to include 
 coherence effects~\cite{mc98} from multi-gluon emission 
 across large rapidity intervals, not associated with 
 small angles.

The results for the matrix elements  in exclusive form are given by  
\begin{equation}
\mathcal{M}_{qg \to qg} =   C_1  {\cal A}_{1}^{(ab)}  + 
   {\overline C}_1    {\cal A}_{1}^{(nab)}   \;\;    ,   \;\;\;   
   \mathcal{M}_{gg \to q\overline q}  =C_2  {\cal A}_{2}^{(ab)}  + 
   {\overline C}_2    {\cal A}_{2}^{(nab)} \;\;    ,   \;\;\; 
\mathcal{M}_{gg \to gg}=C_3  {\cal A}_{3}  
\end{equation}
where 
\begin{equation}
\label{a1ab}
{\cal A}_{1}^{(ab)}
=\Bigg(\frac{\,k_1\,k_2}{k_1\,p_2}\Bigg)^2\frac{(k_1\, p_2)^2+(p_2\, p_3)^2}{(k_1 
\, p_4)\,\,(p_3 \, p_4)}     \;\; , 
\end{equation}
\begin{equation}
\label{a1nab}
{\cal A}_{1}^{(nab)}=\Bigg(\frac{k_1\,k_2}{k_1\, 
p_2}\Bigg)^2
\frac{(k_1\, p_2)^2+(p_2\, p_3)^2}{(k_1 \, p_4)\,\,(p_3 \, p_4)}
\left( \frac{(p_3 \, p_4) \,\,(k_1\, p_2)}{  (k_1  \, p_3) \,(p_2 \, p_4)}+\frac{(k_1 \, 
p_4)\,\,(p_2\, p_3) }{   (k_1  \, p_3)  \,(p_2 \, p_4)}-1\right)   \;\; ,  
\end{equation}
\begin{equation}
\label{a2ab}
{\cal A}_{2}^{(ab)}=\Bigg(\frac{k_1\,k_2}{k_1\,p_2}\Bigg)^2\frac{(p_2\, 
p_3)^2+(p_2\, p_4)^2}{(k_1 \, p_4)\,\,(k_1 \, p_3)}  \;\; , 
\end{equation}
\begin{equation}
\label{a2nab}
{\cal A}_{2}^{(nab)}=\Bigg(\frac{k_1\, k_2}{k_1\, 
p_2}\Bigg)^2
\frac{(p_2\, p_3)^2+(p_2\, p_4)^2}{(k_1 \, p_4)\,\,(k_1 \, p_3)}
\left( \frac{(k_1 \, p_4) \,\,(p_2\, p_3)}{(p_3 \, p_4)  \,(k_1 \, p_2)}+\frac{(k_1 \, p_3)\,\,(p_2\, 
p_4) }{(p_3 \, p_4)  \,(k_1 \, p_2)}-1\right)  \;\; ,  
\end{equation}
\begin{equation}
\label{a3nab}
{\cal 
A}_{3}  = \Bigg(\frac{k_1\,k_2}{k_1 \, p_2}\Bigg)^2
\frac{ (p_3\, 
p_4)( k_1\, p_2)\!+\!(k_1\, p_4) (p_2\, p_3)\!+\!(p_2\, p_4) (k_1\, p_3) }
{(p_2\, p_4) (k_1\, p_4) (p_3\, p_4) (k_1\, p_2) 
(p_2\,p_3) (k_1\,p_3)} 
\left[(p_2\, p_4)^4\!+\!(k_1\, p_2)^4\!+\!(p_2\, p_3)^4\right]   \;\; , 
\end{equation}
and  $C_1 = g^4 (N_c^2 -1 ) / (4 N_c^2) $,  $ {\overline C}_1  = C_1 C_A / (2 C_F)$, 
$C_2 = g^4  / (2 N_c) $,  
$ {\overline C}_2  = C_2 C_A / (2 C_F)$, $C_3 = g^4  N_c^2 / ( N_c^2 -1) $.

The  results above contain the 
 dependence  on the transverse  momentum  k$_\perp$  along the parton lines 
that   connect the hard scatter to the parton distributions.  Nevertheless,  they are 
  short-distance    in the sense that they can be   safely  integrated 
   down to  k$_{\perp} = 0$.  That is, the high-energy 
   projection is   designed so that    all infrared contributions are   factored out in the 
   nonperturbative  Green's functions  $\phi$ in     Eq.~(\ref{forwsigma}). 
An explicit numerical illustration  
 is given in the next section.\footnote{Although 
the  hard-scattering functions  constructed from the amplitudes in 
Eqs.~(\ref{a1ab})-(\ref{a3nab})
     are  not       coefficient functions in the conventional sense of the 
operator product expansion,  
they can be related to  such objects, for inclusive variables, 
along the lines e.g. of~\cite{ch94}.  
 They  could  be    interpreted in terms of    
  coefficient functions in the sense of the high-energy 
OPE of~\cite{balit}.}  

The  role    of 
Eqs.~(\ref{a1ab})-(\ref{a3nab})  is     twofold. 
On one hand,  they give the  high-energy  limit of multi-parton 
matrix elements in the forward region,  which  may  be of direct 
phenomenological significance.  On the other hand, because of the 
factorization theorem~\cite{hef}, logarithmically enhanced 
corrections   for large rapidity     can be 
 systematically obtained to all orders  in $\alpha_s$  from those in 
 the (unintegrated)  distributions for parton splitting  once   the   
  hard scattering functions are  known 
    at   finite    $ {\rm{k}}_\perp$.  To this end the detailed form of the fall-off at large 
    $|  {\rm{k}}_\perp |^2$ is relevant. 

In the next section we  discuss 
 the  behavior at  high transverse momentum   numerically.  
This behavior  reflects 
  properties  of  gluon emission  at large angle   encoded 
  in the  high-energy amplitudes.   
These are relevant,    along with 
   large-angle  effects  in the Sudakov region 
   (see e.g.~\cite{dokmar}),   
   to achieve a full treatment of  
 gluon   coherence  effects~\cite{coh80} capable of describing jet final  
 states  across the whole  rapidity phase space.  
 A uniform treatment of the  high-energy and Sudakov regions     
 is still an open issue~\cite{andetal06} of interest   
for parton-shower  implementations.  

\section{Numerical results }
\label{sec4}

   We now         partially  integrate   the 
 amplitudes   over final states. 
We work at the level of hard scattering matrix elements, leaving  the treatment of 
 parton   
evolution   by showering  to a separate study~\cite{prepar}.   We concentrate on the  
region of hard emissions, where jets are well separated.  
Regions near the boundary of the angular 
 phase  space  are   sensitive to  infrared radiation and  can be addressed 
within a full parton-shower description of the  process.

\begin{figure}[htb]
\vspace{60mm}
\includegraphics{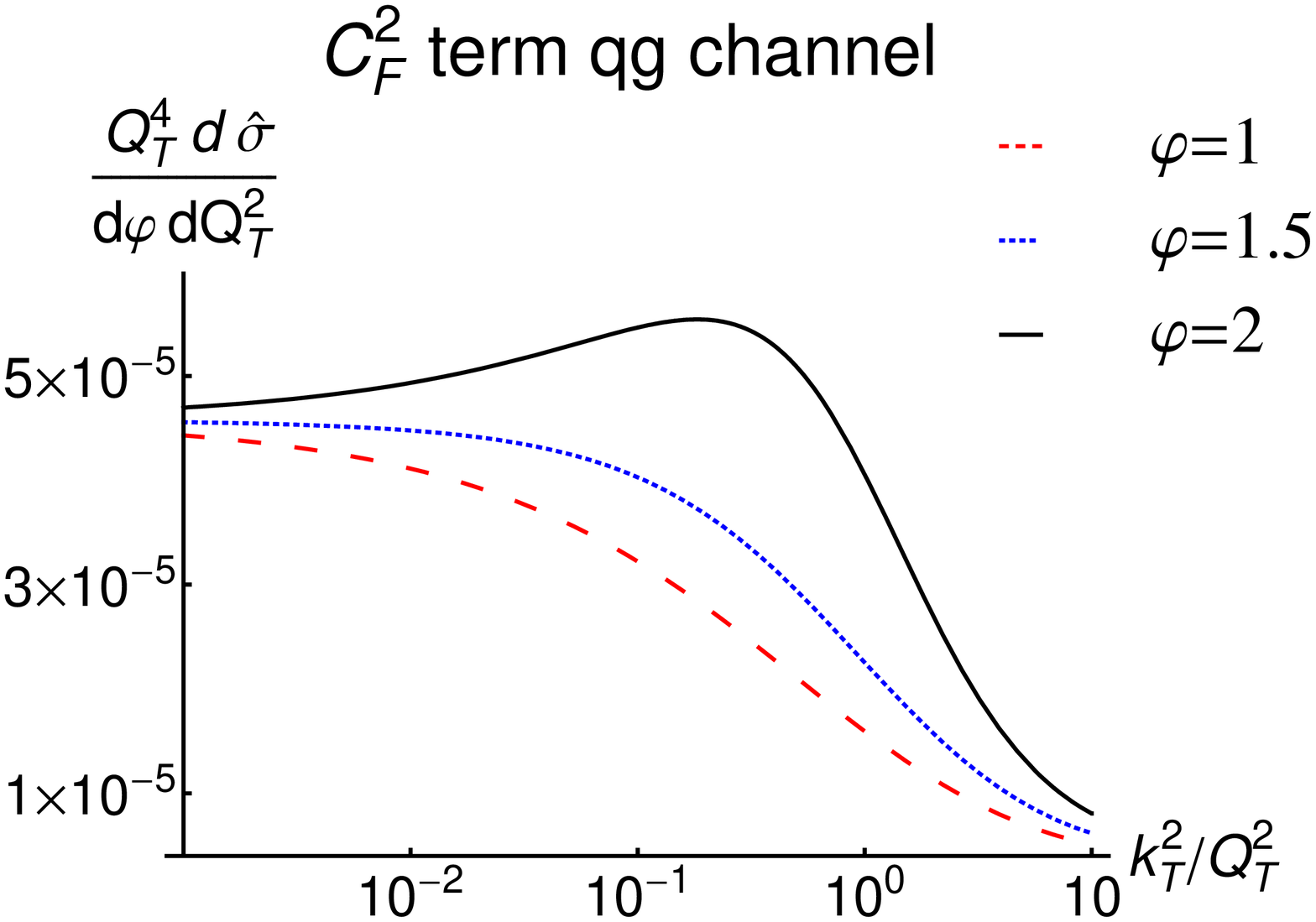}
\includegraphics{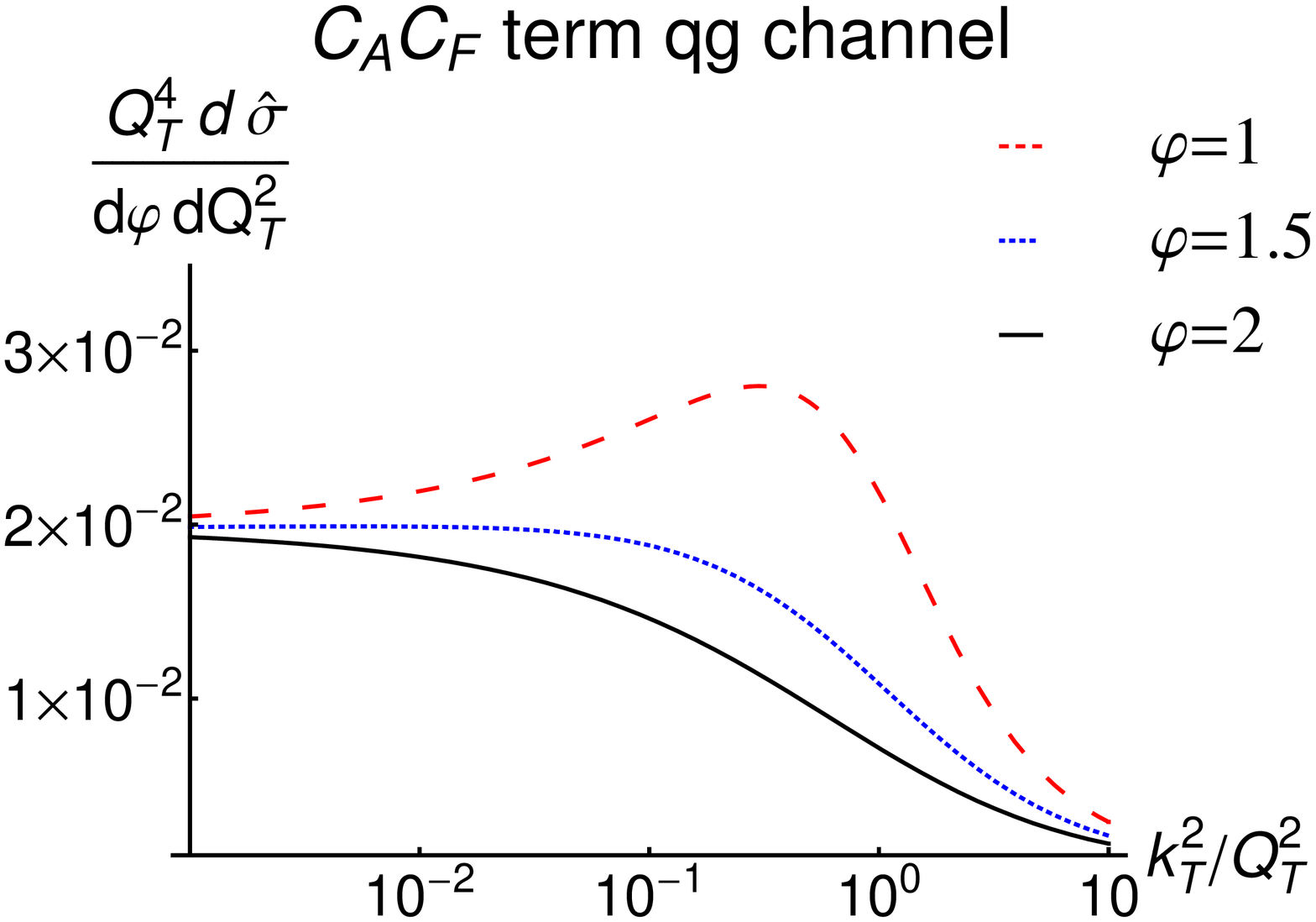}
\caption{The  $k_T / Q_T$  dependence of the  
factorizing   $q g$   hard  cross 
section at high energy:           
 (left) $C_F^2$ term;  (right) $C_F   
C_A$ term  ($\xi_1 \xi_2 S /  Q_T^2= 10^2$,  $\alpha_s = 0.2$).} 
\label{fig:forwplot}
\end{figure}

\begin{figure}[htb]
\vspace{60mm}
\includegraphics{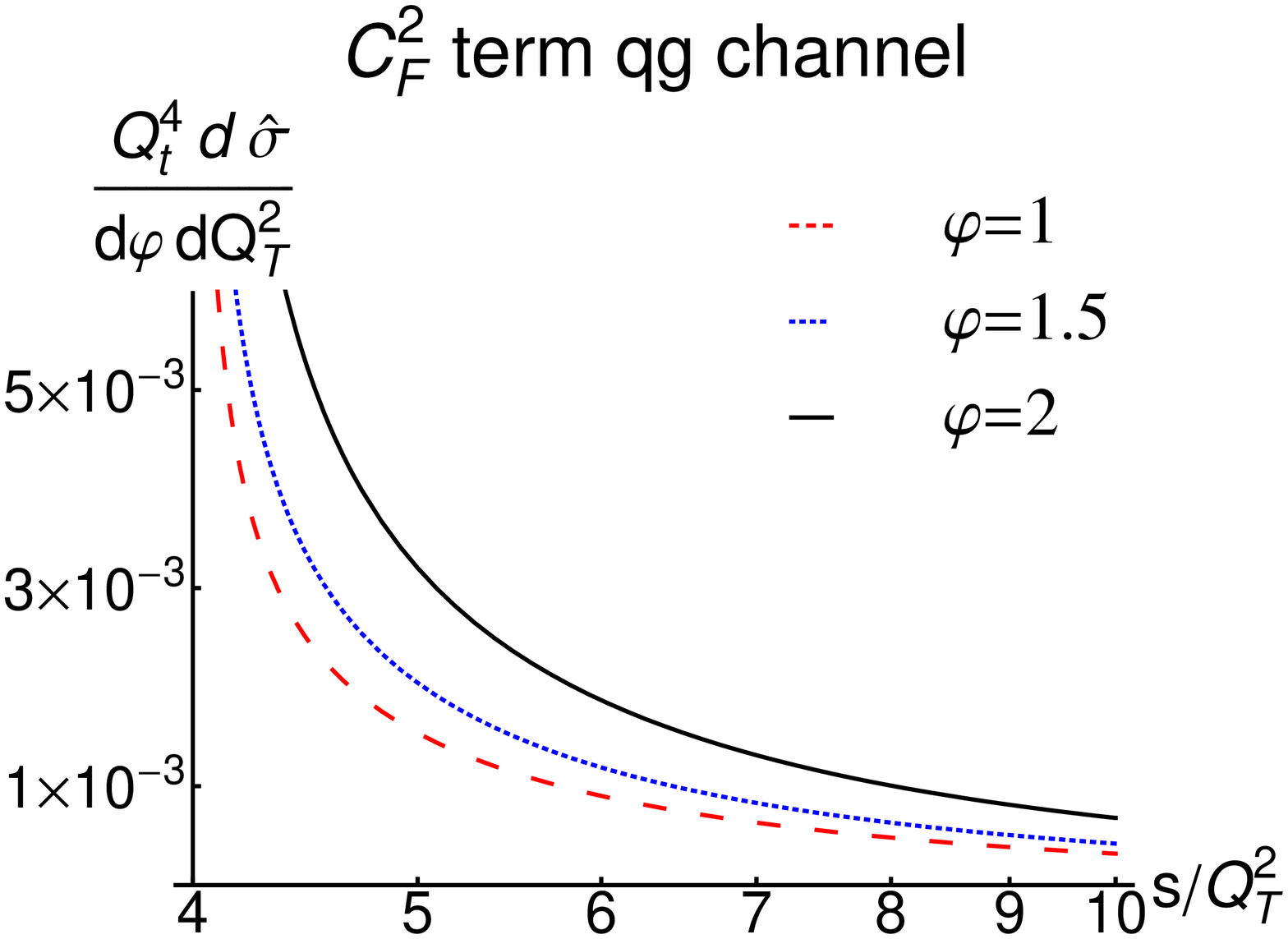}
\includegraphics{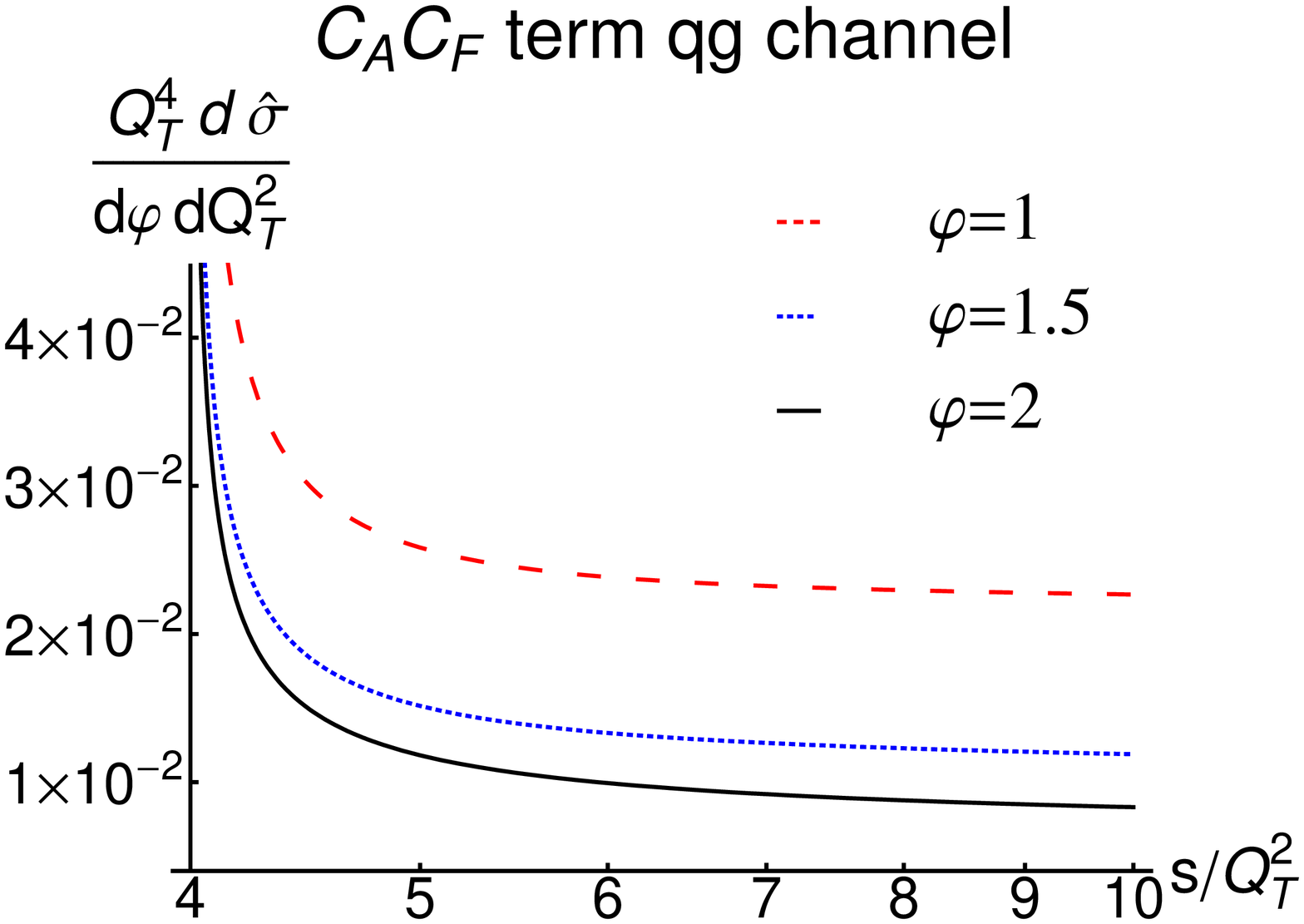}
\caption{The  energy dependence of the    $q g$ hard cross section ($  k_T / Q_T =1$).} 
\label{fig:forwplot1}
\end{figure}

We consider the differential 
distribution   in  the transverse variable $Q_T$ and  azimuthal angle  $\varphi$.  
  The variable  $Q_T$  describes the  
imbalance  in transverse momentum 
between the   hardest  jets, weighted by $\nu$,  according to Eq.~(\ref{qtdef}). 
In Figs.~\ref{fig:forwplot}  and \ref{fig:forwplot1}    we  show 
numerical results   versus transverse momentum and versus 
energy ($q g$ channel). 

The curves in Fig.~\ref{fig:forwplot}   measure   
  the  $k_T$ distribution  
  of   the      jet  system    recoiling against the leading di-jets. 
The result at   $ k_T / Q_T   \to  0$  in these plots returns  the 
 lowest-order result, i.e.,  
  the leading-order process  with two back-to-back 
jets, 
\begin{equation}
\label{born}
{{d   {\widehat  \sigma}   } \over 
{ d Q_T^2 d \varphi  }}  \to   \alpha_s^2 f^{(0)} ( p^2_T / s ) \;\; , \;\;\; 
Q_T \to p_T = | p_{T  3} | =  | p_{T  4} |      \;\; ,
\end{equation}
where $s = (p_3+ p_4)^2$, and  $f^{(0)}$ is given by 
\begin{equation}
\label{f0}
 f^{(0)} (z) = 
{1 \over  
{ 16  \sqrt {1- 4 z} } } \  \left[ C_F^2 z ( 1+ z ) + 2 C_F C_A ( 1 - 3 z  + 
z^2  ) \right]    \;\; . 
\end{equation}
   The dependence on   $ k_T   $ and $\varphi$   plotted 
in Fig.~\ref{fig:forwplot}     is the result of 
higher-order gluon radiation,   treated according 
to  the high-energy asymptotics.  The different behaviors in 
 $\varphi$  for the  $C_F^2$  and $C_A C_F$  terms  reflect the fact 
 that  the former   comes from   the insertion of 
  gluons  on fermion-exchange amplitude 
 while  the latter 
  comes from   the insertion of 
  gluons  on vector-exchange amplitude.  

Fig.~\ref{fig:forwplot1}   shows the energy dependence  for fixed $  k_T / Q_T$. 
The constant asymptotic behavior at large $s$ due to  
 color-octet spin-1 exchange  distinguishes 
the $C_F C_A$ term from the $C_F^2$ term. 
The   dependence on the azimuthal angle 
in    Figs.~\ref{fig:forwplot}  and \ref{fig:forwplot1}  
is relevant,  especially because   forward  jet    measurements  will rely 
  on azimuthal plane correlations between  jets  
far apart  in rapidity (Fig.\ref{fig:azimcorr}).

While Eq.~(\ref{f0})  gives the  collinear emission limit, 
we see     from   Fig.~\ref{fig:forwplot}      that  
multi-gluon radiation  at finite angles  sets a dynamical cut-off 
 at values of  $ k_T$  of order $Q_T$, 
\begin{equation}
\label{mucutoff}
k_T   \lesssim      \mu = c   \   Q_T  \;\; . 
\end{equation}   
The  physical meaning of this  
result    is that the 
  summation  of  the higher-order  logarithmic corrections 
 for large  $y \sim  \ln s / p_T^2$  is precisely  
 determined~\cite{hef,mc98}    by  convoluting the 
 unintegrated splitting 
 functions   over the  $k_T$-dependence  in  Fig.~\ref{fig:forwplot}, 
 via the distributional relation   
\begin{equation}
\label{distrib}
\int     \  d^2 k_T \    \left( {1 \over  k_T^2 } \right)_+    \    {\widehat  \sigma}  
(  k_T ) =   \int        \  d^2 k_T  \    {1 \over  k_T^2 } \ 
\left[   {\widehat  \sigma}  
(  k_T )  -   \Theta (\mu -  k_T ) \  {\widehat  \sigma}  
(  0_T ) \right]  
  \;\; . 
\end{equation}  
So the results       in   
Fig.~\ref{fig:forwplot}  
 illustrate    quantitatively  the  significance of  
contributions  
with   $k_T \simeq  Q_T$ in the large-$y$   region.    
  Non-negligible 
  effects arise at high energy    from the finite-$k_T $  tail. 
These  effects are not included in  collinear-branching  
 generators  (and only partially in fixed-order 
 perturbative calculations),   
 and  become  more and more important  as  the jets are  observed at 
large rapidity separations.

Observe that calculations  based on the unintegrated formalism will in 
general   depend on two  
scales, $\mu$ and the rapidity,  or angle,   cut-off~\cite{jcc-lc08,andetal06,jcc01}.   See 
e.g. the one-loop calculation~\cite{fhfeb07}  for  an explicit example.    
It will be of interest to investigate the   effect  of      Eq.~(\ref{mucutoff})   
on the behavior  in  rapidity distributions~\cite{prepar}. 

\begin{figure}[htb]
\vspace{47mm}
\includegraphics{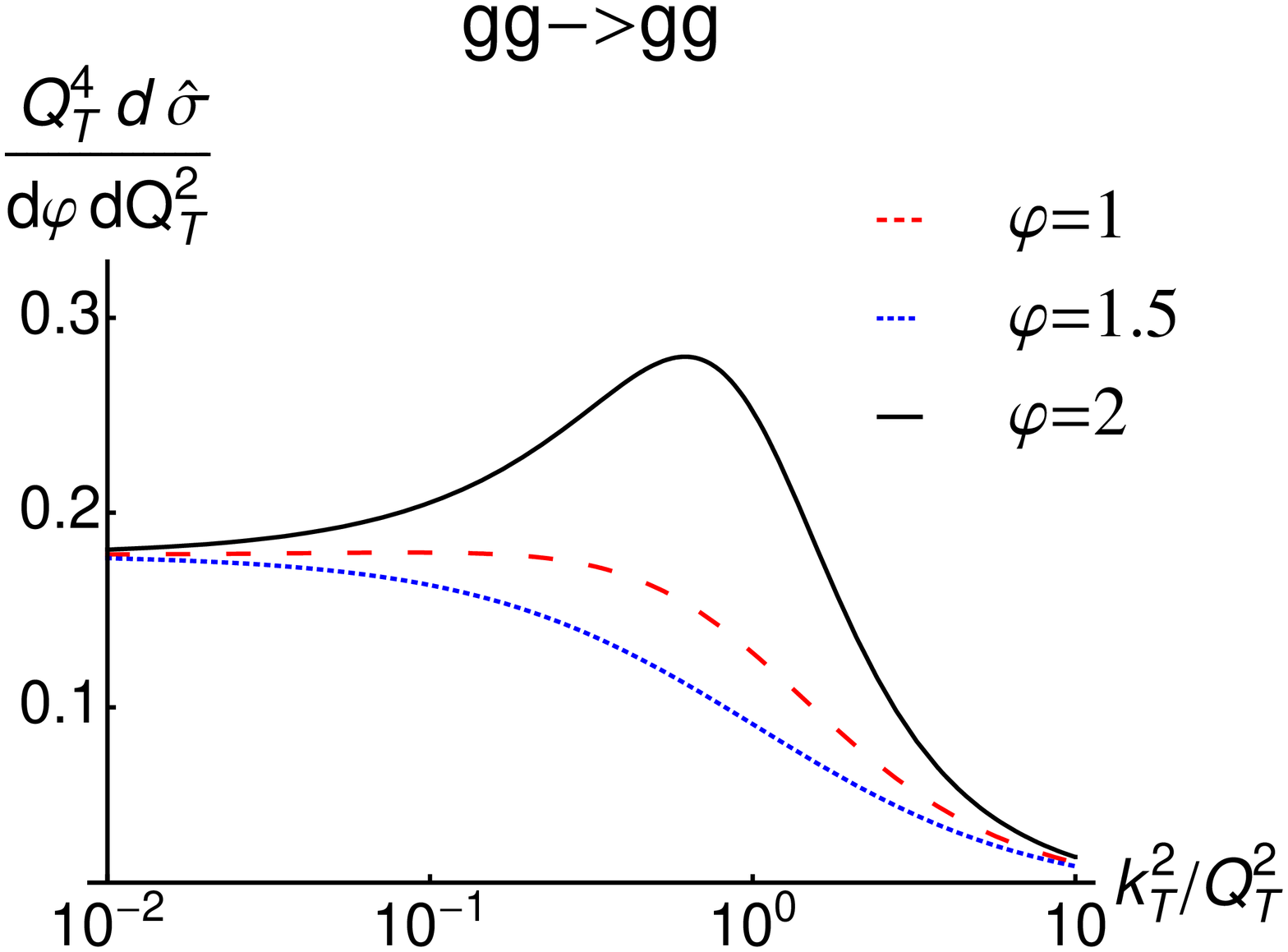}
\includegraphics{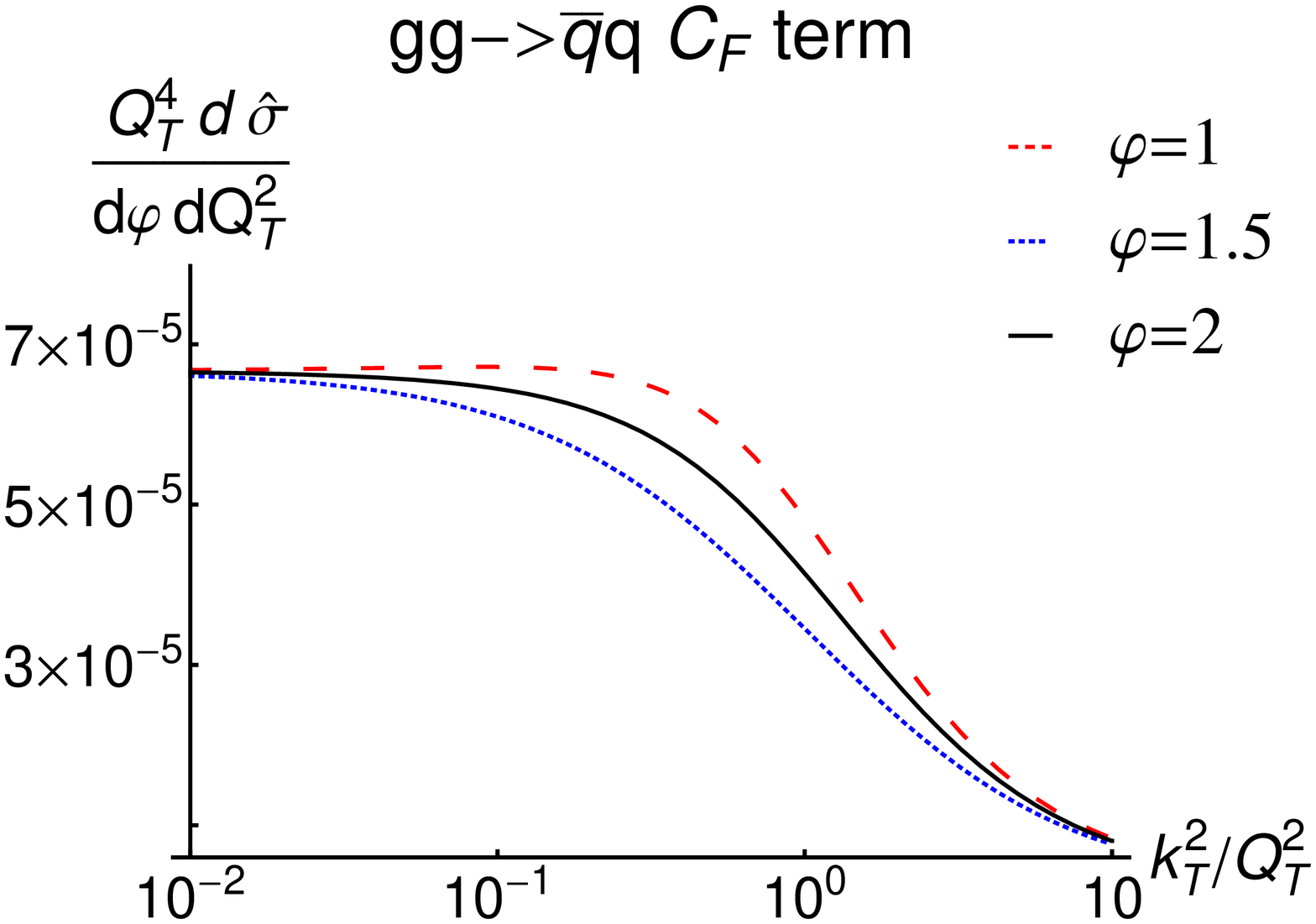}
\includegraphics{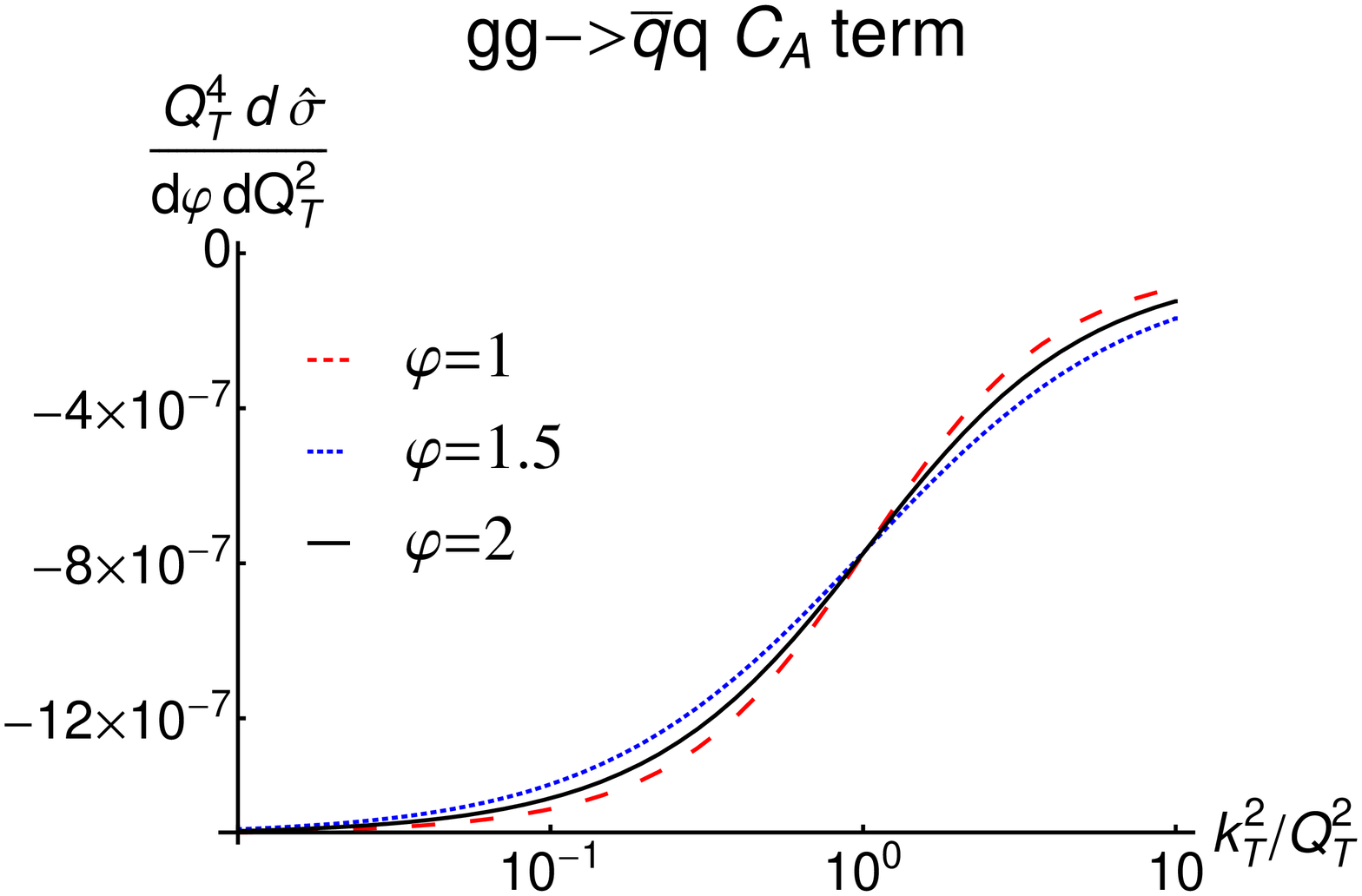}
\caption{The  $k_T / Q_T$  dependence of the  
factorizing   $g g$   matrix elements:           
 (a) $gg \to gg$;  (b) $gg \to q {\bar q} $    $C_F$ term; 
(c) $gg \to q {\bar q} $ $C_A$  
term ($\xi_1 \xi_2 S /  Q_T^2= 10^2$,  $\alpha_s = 0.2$).} 
\label{fig:gg}
\end{figure}

Results  
for  gluon-gluon channels 
 are reported in Fig.~\ref{fig:gg}.   
 Note the large effect  of the purely gluonic component. 
The behavior  versus  $k_T $  is qualitatively  similar  to that in 
 Fig.~\ref{fig:forwplot}.   
Calculations   in progress~\cite{prepar},   including parton showering,  
indicate that     quark 
  and gluon  channels give contributions   of comparable size 
 in  the    LHC forward kinematics. The inclusion of both     is   relevant  for  
 realistic 
 studies of  phenomenology~\cite{cerci,mn_pheno}.  
Since the forward kinematics selects 
asymmetric parton momentum fractions,   effects   
     due to    the   $ x \to 1$   endpoint   behavior~\cite{fhfeb07,jccfh}   
  at  fixed transverse momentum  may become  phenomenologically 
  significant as well.

We conclude this section by  recalling that 
dynamical effects  of high parton densities  have been 
studied~\cite{denterria,ianmue} 
as potential contributions to 
 forward jet events.  We note that  if   
 such  effects  show up at the LHC, 
 the unintegrated   formulation   discussed above 
 would likely be  the   natural  framework  
 to implement   this  dynamics  at   parton-shower level.

\section{Conclusions}
\label{sec:concl}

Forward + central detectors  at the LHC  
 allow  
jet   correlations   to be  measured 
across  rapidity intervals of  several  units,   
 $\Delta y   \greatersim 4 \div 6$.      
   Such multi-jet    states   can   be relevant to 
   new particle  discovery processes as well as 
    new aspects   of  standard model physics.  
    
Existing sets of  forward-jet  data in ep collisions, much more limited  
    than the potential LHC  yield, 
  indicate  that neither conventional parton-showering Monte Carlo generators 
nor next-to-leading-order   QCD calculations  are capable of  describing 
  forward jet    phenomenology.  
 These observations motivate studies of 
  improved methods to    compute    QCD   predictions 
  in the  multiple-scale kinematics   implied by the forward region.  

We have analyzed the  high-energy factorization  
  that 
serves  to  sum 
consistently to  higher orders in $\alpha_s$ 
 both the  logarithmic corrections  
 in the large rapidity 
interval and those   in the hard jet transverse energy.  
We have determined the 
gauge-invariant  (though not on shell)  
  high-energy  amplitudes, which are needed to evaluate the factorization formula   
     for    forward  jet   hadroproduction. 

  Our results  can  
     be used   along with    k$_\perp$-dependent  kernels for parton branching.  
      They can  
    serve to  construct predictions for   exclusive  observables   associated 
     to forward jets,  including jet correlations, 
     that  take  into account  gluon   coherence not only in the collinear emission region but 
     also in the large-angle emission region.

 \vskip 1 cm

\noindent    {\bf Acknowledgments}. 
One of us  (F.H.)  visited  DESY  at various stages 
  while  this work was being done and  wishes to 
  thank   the DESY directorate   for hospitality and support. 
Part of this work was  carried out   during  
the 2009 DESY Institute on 
Parton Showers and Resummations. 
We thank the organizers and 
participants in the workshop for  the stimulating 
atmosphere.  We gratefully acknowledge   useful  discussions  
with  S.~Baranov, J.~Collins,   A.~Knutsson, 
A.~Lipatov, Z.~Nagy,  T.~Rogers   and   N.~Zotov.

\end{document}